\documentclass{article}
\usepackage{spconf,amsmath,graphicx}
\usepackage{hyperref}
\usepackage{enumitem}
\usepackage{amssymb}
\usepackage{xcolor}

\newcommand{\fzero}{F\textsubscript{0}}

\title{Controllable Prosody Generation With Partial Inputs}

\name{
    \parbox{\linewidth}{
        \centering Dan Andrei Iliescu$^{\star}$ \qquad Devang S Ram Mohan$^{\dagger}$ \qquad Tian Huey Teh$^{\ddagger}$ \qquad   Zack Hodari$^{\dagger}$ \thanks{$^{\star}$Work completed while at Papercup Ltd.}}
    }
\address{
    $^{\dagger}$ Papercup Technologies \qquad
    $^{\star}$ University of Cambridge 
    \qquad
    $^\ddagger$ Google Deepmind\\
    \fontsize{9}{9}\selectfont\ttfamily\upshape
	zack@papercup.com \qquad dai24@cam.ac.uk \\
}

\begin{document}
%
\maketitle
\begin{abstract}
We address the problem of human-in-the-loop control for generating prosody in the context of text-to-speech synthesis.
Controlling prosody is challenging because existing generative models lack an efficient interface through which users can modify the output quickly and precisely.
To solve this, we introduce a novel framework whereby the user provides partial inputs and the generative model generates the missing features.
We propose a model that is specifically designed to encode partial prosodic features and output complete audio. 
We show empirically that our model displays two essential qualities of a human-in-the-loop control mechanism: efficiency and robustness. 
With even a very small number of input values (\raisebox{0.5ex}{\texttildelow}4), our model enables users to improve the quality of the output significantly in terms of listener preference (4:1).
\end{abstract}
\section{Introduction}
\label{intro}

Human-in-the-loop (HitL) control involves the interplay between human expertise and machine learning. This is crucial in numerous applications, particularly those generating high-dimensional data like text, speech, or images. Deep learning models are able to learn statistics from data that can help users turn their intuition into realistic outputs~\cite{wu2022survey}.

Achieving fine-grained control is challenging due to the complexity of high-dimensional data. Existing methods range from fine-grained to coarse-grained. Fine-grained methods, like in-painting for image generation, allow for localized adjustments, while coarse-grained approaches like language model prompting affect more global features. Generative models for text-to-speech (TTS) struggle to offer practical methods for fine-grained control. Users must either manually specify all the conditioning inputs \cite{Lee2018RobustAF, vetterli}, which is costly and error-prone, or provide high-level conditioning labels that are too general~\cite{Ramesh2022HierarchicalTI}.

In particular, we are interested in controlling the prosody of generated speech. Prosody is the component of human speech that conveys emotion, attitude, intentions, and other communicative functions~\cite[Chapter~2]{zack-PhD-thesis:2022}. Prosody is communicated through changes in intonation, loudness, phrasing, timing, and voice quality~\cite{prosody-review:2010}. It is common to approximate these perceptual factors at the phoneme-level using prosodic acoustic features (PAFs): \textit{\fzero}, \textit{energy}, and \textit{duration}~\cite{PaperCup-Ctrl-P:2021}. Our goal is to design a generative model that allows the user to choose the rendition they desire. This is is an important goal as prosody is inherently one-to-many; the same sentence can be spoken in many different ways~\cite{prosody-review:2010}.


\begin{figure}[t]
    \begin{center}
        \centerline{\includegraphics[width=0.9\columnwidth]{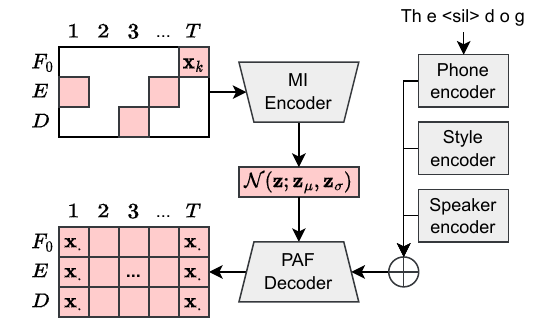}}
        \caption{Our model, MICVAE, encodes partial information and decodes a complete output. The inputs and outputs are prosodic acoustic features (PAFs), with 3 values (\fzero, energy and duration) for each phoneme in the sentence.}
        \label{fig:model}
    \end{center}
    \vskip -0.4in
\end{figure}

In this work, we claim that partial inputs from users can serve as an effective method for fine-grained prosody control. The user provides values for a subset of PAFs, called control points, and the model fills in the rest, called missing points. This allows for localised and flexible adjustments without overwhelming the user.

We design our own model for prosody generation and control, called Multiple-Instance Conditional Variational Autoencoder (MICVAE). MICVAE can be conditioned on partially specified inputs via its novel self-attention encoder, which treats the individual PAFs as instances in an unordered set.

We show empirically that our model is an effective means of HitL control. We evaluate two key requirements for user-friendly control:
1) \textbf{Efficiency:} Achieving the desired output should require the fewest number of control points.
2) \textbf{Robustness:} The model's outputs should be realistic regardless of the pattern in which the control points are selected. We present the following contributions:

\begin{enumerate}[itemsep=0pt,parsep=0pt,partopsep=0pt]
    \item We propose MICVAE, a novel probabilistic generative model for prosody control that is robust to differing patterns of missingness.
    \item We demonstrate that using partial inputs to control prosody generation is efficient and robust.
    \item We introduce iterative refinement, a novel automated evaluation procedure designed to simulate HitL control in a reproducible way.
\end{enumerate}


\section{Related Work}

\textbf{\textsc{Human-in-the-loop Control}.} Research on human-in-the-loop machine learning is growing rapidly \cite{wu2022survey}.
Unfortunately, there are few works where humans control the outputs of deep models at test time. Correcting high-dimensional data manually is prohibitively costly. In the context of Computer Vision, some approaches have been studied extensively, such as inpainting \cite{Elharrouss2019ImageIA} and semantic manipulation \cite{park2019SPADE}. However, even these methods struggle with generating globally consistent images \cite{Zhang2023TowardsCI}.

One approach to enable human control is to use the latent representation as input. However, not all representations are amenable for human interaction. Various methods can be applied to encourage interpretability \cite{Tschannen2018RecentAI}. Structure can be imposed on the latent variables through independence assumptions \cite{prosody-control:VAE-MI:2021} and hierarchical assumptions \cite{Hsu2017UnsupervisedLO}. Self-supervised losses can also be used to disentangle concepts \cite{prosody-control:VAE-3dim:2020}. Discrete latent variables are another useful method to make representations more usable for humans \cite{VQ-VAE:2017}.


\textbf{\textsc{Prosody}.}
Prosody prediction is a challenging task because models lack the relevant context information that humans use to plan prosodic choices \cite{context-definition:1992}. Furthermore, even if we were to provide additional contextual information, there may be no single best prosodic rendition for a given situation \cite{zack-SSW19:2019}. Control of prosody is complicated by the many communicative purposes that prosody serves \cite{grounding:1991}



\textbf{\textsc{Inpainting}.} The closest related work to our own is that on speech inpainting \cite{Borsos2022SpeechPainterTS}, itself drawing upon the vast literature on image inpainting \cite{Elharrouss2019ImageIA}. 
However, our work explicitly addresses the problem of control. Our introduction of new criteria for success (efficiency and robustness) differentiates us from inpainting. For example, methods in our framework must produce consistent prosody across various missingness patterns, while inpainting evaluates models under specific patterns (e.g., a contiguous area of the image).

\section{Multiple-Instance CVAE}

We introduce a specialized Conditional Variational Autoencoder (CVAE) \cite{conditional-VAE:2015} called Multiple Instance CVAE (MICVAE). as our acoustic feature predictor (AFP) model. The overarching architecture is illustrated in Figure~\ref{fig:model}. In MICVAE, user-defined control points (individual PAF values) are the encoder inputs. The encoder's latent representation is decoded conditioned on phones to generate a full sequence of PAFs, these are then used to synthesise audio. The conditioning labels are the content embeddings (i.e. phoneme, style and speaker embeddings).


\subsection{Multiple-Instance Encoder}

\begin{figure}[t]
    \begin{center}
        \centerline{\includegraphics[width=\columnwidth]{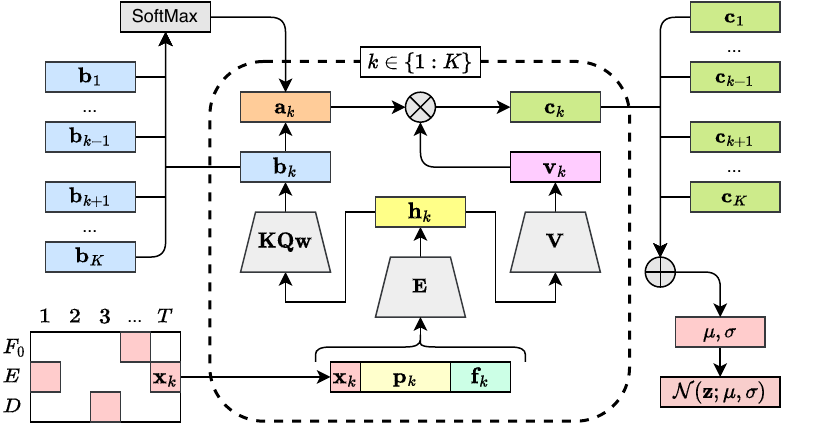}}
        \caption{Our novel ``multiple-instance'' encoder. The control points of the partial input are treated as an unordered bag of features. They are aggregated into a fixed-length vector by a self-attention mechanism.}
        \label{fig:encoder}
    \end{center}
    \vskip -0.4in
\end{figure}


MICVAE's encoder is the key design element that enables users to provide partial inputs to control prosody. As shown in Figure~\ref{fig:encoder}, the encoder aggregates user inputs into a latent representation, denoted as \(\mathbf{z}\), using self-attention. The encoder's design is similar to self-attention aggregators used for multiple-instance learning \cite{Ilse2018AttentionbasedDM}. We now explain Figure~\ref{fig:encoder}.

\textbf{Input Preprocessing.} The encoding process starts with each control PAF, \(\mathbf{x}_k \in \mathbb{R}\), where \(k\) indexes these control PAFs over a set \(\{1:K\}\). The positional encodings \(\mathbf{p}_k \in \mathbb{R}^P\) and feature stream encodings \(\mathbf{f}_k \in \mathbb{R}^F\) are concatenated with \(\mathbf{x}_k\), to produce an extended feature vector \([ \mathbf{x}_k, \mathbf{p}_k, \mathbf{f}_k ]\). The positional encodings follow a sinusoidal pattern, while the feature encodings consist of learned vectors. Hidden representations are computed as: $\mathbf{h}_k = \mathrm{ReLU} \left( \mathbf{E} ~ [ \mathbf{x}_k, \mathbf{p}_k, \mathbf{f}_k ]^T \right)$.

\textbf{Value Vector Computation.} Each intermediate representation \(\mathbf{h}_k \in \mathbb{R}^H\) is then transformed into a value vector, \(\mathbf{v}_k \in \mathbb{R}^D\), via a perceptron: $\mathbf{v}_k = \tanh \left( \mathbf{V} \mathbf{h}_k^T \right)$.

\textbf{Attention Weight Calculation.} The attention weights \(\mathbf{a}_k \in [0, 1]\) are derived from key-query products \(\mathbf{b}_k\). Similar to \cite{Ilse2018AttentionbasedDM}, this is what allows the encoder to be invariant to the number of user inputs: $\mathbf{b}_k = \mathbf{w}^T \left( \tanh \left( \mathbf{Q} \mathbf{h}_k^T \right) \odot \mathrm{sigm} \left( \mathbf{K} \mathbf{h}_k^T \right) \right)$, $\mathbf{a}_k = \exp \left( \mathbf{b}_k \right) / \sum_{l=1}^{K} \exp \left( \mathbf{b}_l \right)$.

\textbf{Aggregation into Latent Embeddings.} These attention weights are used to compute a weighted sum of value vectors, yielding a sentence-level embedding \(\mathbf{z}^\prime \in \mathbb{R}^D\): $\mathbf{z}^\prime = \sum_{k=1}^{K} \mathbf{a}_k \mathbf{v}_k$.

\textbf{Latent Variable Parameterization.} Finally, this embedding \(\mathbf{z}^\prime\) is projected to parameterize a Gaussian latent variable \(\mathbf{z}\): $
\mathbf{z} = \mathcal{N} \left( \mu, \sigma \right), \quad \mu = \mathbf{U} \mathbf{z}^{\prime T}, \quad \sigma = \mathbf{S} \mathbf{z}^{\prime T}$.

\textbf{Parameters.} The model includes the following trainable parameters: $\mathbf{K}$, $\mathbf{Q} \in \mathbb{R}^{L \times H}$, $\mathbf{w} \in \mathbb{R}^{D \times L}$, $\mathbf{V} \in \mathbb{R}^{D \times H}$, $\mathbf{E} \in \mathbb{R}^{H \times (1+P+F)}$, and $\mathbf{f} \in \mathbb{R}^{F \times 3}$. The dimensions of the various components are as follows: \( H = 64, D = 32, L = 64, F = 8, P = 8 \).

In addition to the Multiple-Instance encoder used to enable control, our model uses three other encoders to capture the content that is being generated: phonetic content, speaker identity, and style. 
Our PAF decoder has the same architecture as the acoustic feature predictor from \cite{PaperCup-Ctrl-P:2021}. 


\section{Evaluation}

\begin{figure}
    \centerline{\includegraphics[width=\columnwidth]{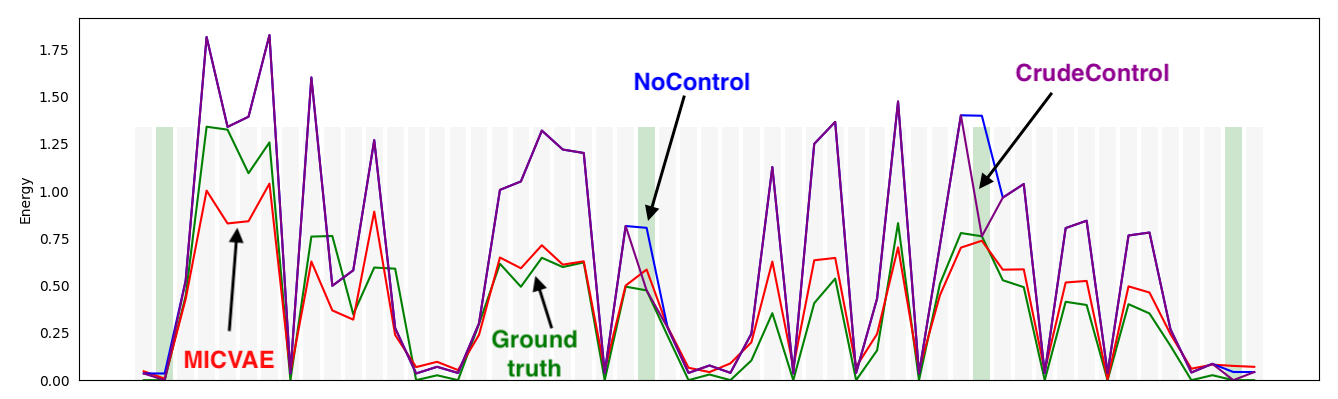}}
    \vskip -0.1in
    \caption{Our model produces plausible prosody, whereas Crude Control produces inconsistent prosody. The control points are shown with green vertical bars.} 
    \label{fig:demo}
    \vskip -5pt
\end{figure}


We evaluate MICVAE's efficiency and robustness using a purpose-built prosody control dataset: a proprietary Latin American Spanish corpus featuring 32 speakers (17 female, 15 male) and comprising approximately 38 hours of expressive speech across ~26,000 utterances. We reserve 800 utterances for validation and 182 for testing, covering a variety of emotional and speaking styles. Data preparation involves one-hot encoding of phonetic, punctuation, and boundary tokens. Pre-emphasis is applied to waveforms, and mel-spectrograms are extracted using 128 bins, a 50ms frame length, and a 10ms frame shift. \(f_0\) is extracted using RAPT \cite{RAPT:1995}, and energy is calculated as the frame-wise root-mean-square of the waveform. Durations are obtained via a Kaldi forced aligner \cite{PoveyASRU2011}, and all features are mean-variance normalized per speaker.

Speech is synthesized using a TTS system conditioned on MICVAE's PAFs. The system includes a Latin American Spanish front-end for text normalization and grapheme-to-phoneme conversion. The acoustic model relies on Ctrl-P \cite{PaperCup-Ctrl-P:2021}, an attention-based Tacotron-2 variant, while the neural vocoder is WaveRNN \cite{waveRNN:2018}. 

Human-in-the-loop (HitL) control is simulated by selecting a subset of PAFs, referred to as control points, from one actor's rendition of a given test sentence. These control points condition our Acoustic Feature Predictor (AFP) model, with the remaining PAFs being generated by the model (Figure~\ref{fig:demo}). We use Root Mean Square Error (RMSE) to quantify the control between model-generated and ground-truth PAFs.




\subsection{Baselines}

\textbf{\textsc{Masked CVAE}.} Masked CVAE is a baseline system that uses the standard approach of masking for missing data imputation and inpainting~\cite{Elharrouss2019ImageIA}. Masked CVAE has a similar architecture to MICVAE, the primary difference lies in the encoder design. Masked CVAE employs a multi-layer Recurrent Neural Network (RNN) as its encoder. Input PAFs are augmented with an additional binary feature for each of the three streams (\fzero, energy, duration) to indicate whether a given PAF is provided (1) or missing (0). Unlike MICVAE, which is inherently robust to varying levels of input missingness, Masked CVAE must be trained with sparsity in order to learn what the values 0,1 in the mask denote. During training, a fixed missingness percentage \( P \) is set, and \( P \% \) of the PAFs are randomly masked out for each training sentence.


\textbf{\textsc{No Control}.} The default prosody model: predicts PAFs from phones, speaker identity, and style code. Uses the same decoder architecture as the MICVAE.

\textbf{\textsc{Crude Control}.} A na\"ive system that forcefully modifies predicted values individually, this straw-man represents a controllable model that is entirely inconsistent. Its default prosody is generated with \textsc{NoControl}, and then modified manually with the control points\footnote{Samples demonstrating TTS for these systems can be found here, \href{https://anonymous-submission-563098.github.io/sparse-control/}{anonymous-submission-563098.github.io/sparse-control}}.

\subsection{Evaluation Results}

\begin{figure}[t]
    \centerline{\includegraphics[width=\columnwidth]{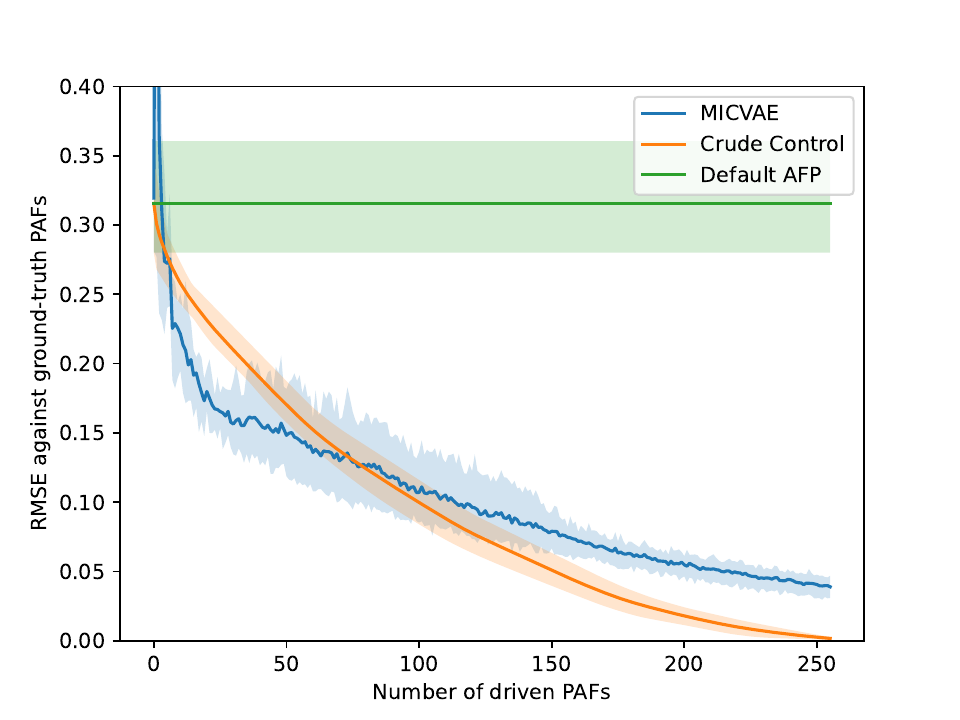}}
    \vskip -0.15in
    \caption{Our model controls prosody efficiently. Its generated PAFs are closer to the ground-truth rendition than  is the manually modified output of the AFP, called \sc{CrudeControl}.}
    \label{fig:iterative}
    \vskip -0.2in
\end{figure}

\begin{figure}[t]
    \begin{center}
        \centerline{\includegraphics[width=0.9\columnwidth]{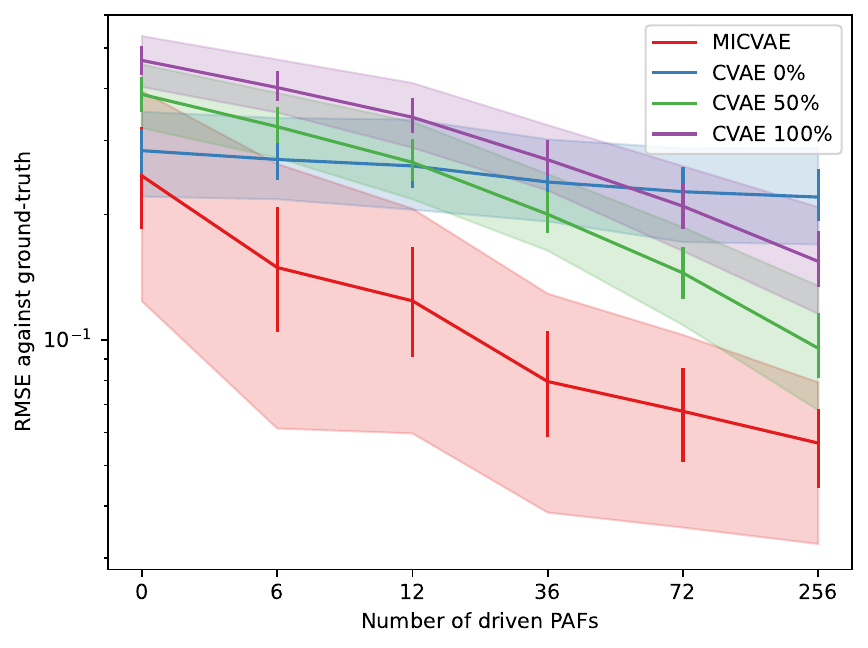}}
        \vskip -0.15in
        \caption{Our model is robust to changes in the missingness pattern, whereas Masked CVAE only performs well when tested on the same missingness pattern as the one on which it was trained.}\label{fig:robustness}
    \end{center}
    \vskip -0.3in
\end{figure}

\begin{figure}
    \begin{center}
    \centerline{\includegraphics[width=\columnwidth]{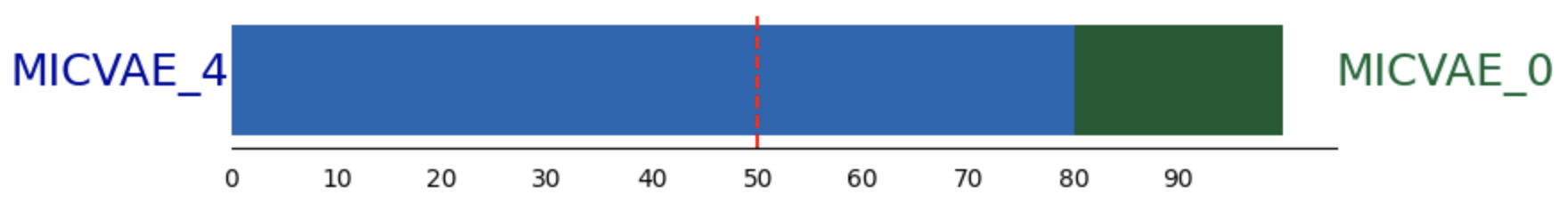}}
    \vskip -0.15in
    \caption{4 control points are enough to bring our model's output significantly closer to the reference utterance. These are ratings in an A/B/R test where the question is ``Which of these is closer to the reference ground-truth control audio?''.}
    \label{fig:faithfulness}
    \end{center}
    \vskip -0.3in
\end{figure}

\textbf{\textsc{Efficiency} (Objective Evaluation).} We assess efficiency by how many control points are required for the model to produce outputs that align with the user's intention. Alignment is measured by the Root Mean Squared Error (RMSE) between the generated and ground-truth PAFs. We feed the model with additional control points from our test set incrementally. In this ``iterative refinement'' process the PAF with the highest RMSE in the previous generation is provided in the next step. A lower RMSE achieved with fewer input PAFs directly translates to efficiency; it implies less user input is required to generate satisfactory prosodic features.

Figure~\ref{fig:iterative} reveals that MICVAE significantly outperforms the \textsc{CrudeControl} method, achieving lower RMSEs across a realistic range of missingness (between 4 and 70 PAFs). Our model (MICVAE) produces prosody closer to the ground-truth than the prosody resulting from manually setting the output PAFs to default values, meaning MICVAE uses the learned structure of prosody data to fill in the gaps between the control points. The fact that Crude Control performs better than MICVAE is to be expected when the missingness rate approaches 0, because we are manually switching all PAFs to the values extracted from the ground-truth utterance.

\textbf{\textsc{Efficiency} (Subjective Evaluation).} We also conduct an A/B/R listening test involving 20 native Latin American Spanish speakers. Participants choose which of two audio outputs (one with 4 input PAFs and one with 0) sounds closer to a reference ground-truth audio. We chose 4 and 0 to demonstrate the impact that just 4 control points can have on the prosody.

Figure~\ref{fig:faithfulness} reveals that MICVAE with 4 input PAFs is significantly perceived as closer to the reference audio compared to the version without any input PAFs. This not only confirms the model's faithfulness to user intentions but also underscores its efficiency, as it achieves these results with a minimum number of input PAFs.

\textbf{\textsc{Robustness} (Objective Evaluation).} A robust model should adapt to different missingness patterns, thereby providing users the flexibility to choose which PAFs to specify. This robustness is essential for accommodating the diversity of user input methods and intents. We examine MICVAE's robustness against varying patterns of missingness by comparing it to Masked CVAE. The evaluation involves running simulated control tests with random selections of control points at different numbers (0, 6, 12, 36, 72, 256).

As illustrated in Figure~\ref{fig:robustness}, MICVAE outperforms all versions of Masked CVAE, even with the same model capacity. The results indicate that Masked CVAEs fail to adapt to varying missingness patterns effectively, emphasizing the robustness of MICVAE. Note that our evaluation results generalise to other generative models than CVAE.\@ We compare different conditioning mechanisms (multiple-instance vs masking) rather than different generator architectures.

\section{Limitations}

We acknowledge that our work does not investigate all plausible strategies for selecting control points. We explored two strategies: randomly selecting a certain number of PAFs or using iterative refinement (sequentially providing all the PAFs in descending order of reconstruction error). However, other strategies exist, including: 1) Providing PAFs for an entire word. 2) Providing a few PAFs at the beginning and the end of a sentence. 3) Providing only F0 values. A detailed user study would evaluate the utility of our control framework in a range of practical settings.

Nevertheless, we believe that our quantitative evaluation already demonstrates the effectiveness of our model as a proof of concept. Our work presents a novel problem (efficiency and robustness) and approach (partial inputs) in order to initiate discussion and collaboration within the community. Future research can expand our work by conducting user studies of how humans choose to manipulate PAFs in practice.

\section{Conclusion}

In this study, we address the challenges of controlling generative models by introducing a new partial-input control framework. We establish three key attributes for a successful generative model: efficiency, robustness, and faithfulness. Our model, derived from the Masked CVAE, empirically exhibits these traits. To enable reproducible testing, we use simulated human-in-the-loop (HitL) control with natural prosodic features extracted from a dataset of repeated utterances. Our work serves as a foundation for future studies on human-in-the-loop control in generative models.


\bibliographystyle{IEEEbib}
\bibliography{main,zack_thesis}

\begin{thebibliography}{10}

\bibitem{wu2022survey}
Xingjiao Wu, Luwei Xiao, Yixuan Sun, Junhang Zhang, Tianlong Ma, and Liang He,
\newblock ``A survey of human-in-the-loop for machine learning,''
\newblock {\em Future Generation Computer Systems}, 2022.

\bibitem{Lee2018RobustAF}
Younggun Lee and Taesu Kim,
\newblock ``Robust and fine-grained prosody control of end-to-end speech
  synthesis,''
\newblock {\em ICASSP 2019 - 2019 IEEE International Conference on Acoustics,
  Speech and Signal Processing (ICASSP)}, pp. 5911--5915, 2018.

\bibitem{vetterli}
Martin Vetterli, Jelena Kovačević, and Vivek~K Goyal,
\newblock {\em Foundations of Signal Processing},
\newblock Cambridge University Press, 2014.

\bibitem{Ramesh2022HierarchicalTI}
Aditya Ramesh, Prafulla Dhariwal, Alex Nichol, Casey Chu, and Mark Chen,
\newblock ``Hierarchical text-conditional image generation with clip latents,''
\newblock {\em ArXiv}, vol. abs/2204.06125, 2022.

\bibitem{zack-PhD-thesis:2022}
Zack Hodari,
\newblock ``Synthesising prosody with insufficient context,''
\newblock 2022.

\bibitem{prosody-review:2010}
Michael Wagner and Duane~G Watson,
\newblock ``Experimental and theoretical advances in prosody: A review,''
\newblock {\em Language and cognitive processes}, vol. 25, no. 7-9, pp.
  905--945, 2010.

\bibitem{PaperCup-Ctrl-P:2021}
Devang S~Ram Mohan, Vivian Hu, Tian~Huey Teh, Alexandra Torresquintero,
  Christopher G.~R. Wallis, Marlene Staib, Lorenzo Foglianti, Jiameng Gao, and
  Simon King,
\newblock ``{Ctrl-P}: Temporal control of prosodic variation for speech
  synthesis,''
\newblock in {\em Proc. Interspeech}, Brno, Czech Republic, 2021.

\bibitem{Elharrouss2019ImageIA}
Omar Elharrouss, Noor Almaadeed, Somaya~Ali Al-Maadeed, and Younes Akbari,
\newblock ``Image inpainting: A review,''
\newblock {\em Neural Processing Letters}, vol. 51, pp. 2007--2028, 2019.

\bibitem{park2019SPADE}
Taesung Park, Ming-Yu Liu, Ting-Chun Wang, and Jun-Yan Zhu,
\newblock ``Semantic image synthesis with spatially-adaptive normalization,''
\newblock in {\em Proceedings of the IEEE Conference on Computer Vision and
  Pattern Recognition}, 2019.

\bibitem{Zhang2023TowardsCI}
Guanhua Zhang, Jiabao Ji, Yang Zhang, Mo~Yu, T.~Jaakkola, and Shiyu Chang,
\newblock ``Towards coherent image inpainting using denoising diffusion
  implicit models,''
\newblock in {\em International Conference on Machine Learning}, 2023.

\bibitem{Tschannen2018RecentAI}
Michael Tschannen, Olivier Bachem, and Mario Lucic,
\newblock ``Recent advances in autoencoder-based representation learning,''
\newblock {\em Third workshop on Bayesian Deep Learning (NeurIPS)}, 2018.

\bibitem{prosody-control:VAE-MI:2021}
Xiaochun An, Frank~K Soong, Shan Yang, and Lei Xie,
\newblock ``Effective and direct control of neural {TTS} prosody by removing
  interactions between different attributes,''
\newblock {\em Neural Networks}, 2021.

\bibitem{Hsu2017UnsupervisedLO}
Wei-Ning Hsu, Yu~Zhang, and James~R. Glass,
\newblock ``Unsupervised learning of disentangled and interpretable
  representations from sequential data,''
\newblock in {\em NIPS}, 2017.

\bibitem{prosody-control:VAE-3dim:2020}
Guangzhi Sun, Yu~Zhang, Ron~J Weiss, Yuan Cao, Heiga Zen, and Yonghui Wu,
\newblock ``Fully-hierarchical fine-grained prosody modeling for interpretable
  speech synthesis,''
\newblock in {\em Proc. International Conference on Speech and Signal
  Processing}. IEEE, 2020, pp. 6264--6268.

\bibitem{VQ-VAE:2017}
Aaron van~den Oord, Oriol Vinyals, and Koray Kavukcuoglu,
\newblock ``Neural discrete representation learning,''
\newblock in {\em Proc. International Conference on Neural Information
  Processing Systems}, Long Beach, USA, 2017, pp. 6306--6315.

\bibitem{context-definition:1992}
Charles~Goodwin et~al.,
\newblock {\em Rethinking context: Language as an interactive phenomenon},
\newblock Cambridge University Press, 1992.

\bibitem{zack-SSW19:2019}
Zack Hodari, Oliver Watts, and Simon King,
\newblock ``Using generative modelling to produce varied intonation for speech
  synthesis,''
\newblock in {\em Proc. Speech Synthesis Workshop}, Vienna, Austria, 2019, pp.
  239--244.

\bibitem{grounding:1991}
Herbert~H Clark and Susan~E Brennan,
\newblock ``Grounding in communication,''
\newblock 1991.

\bibitem{Borsos2022SpeechPainterTS}
Zal{\'a}n Borsos, Matthew Sharifi, and Marco Tagliasacchi,
\newblock ``Speechpainter: Text-conditioned speech inpainting,''
\newblock in {\em Interspeech}, 2022.

\bibitem{conditional-VAE:2015}
Kihyuk Sohn, Honglak Lee, and Xinchen Yan,
\newblock ``Learning structured output representation using deep conditional
  generative models,''
\newblock Montr\'{e}al, Canada, 2015, vol.~28, pp. 3483--3491.

\bibitem{Ilse2018AttentionbasedDM}
Maximilian Ilse, Jakub~M. Tomczak, and Max Welling,
\newblock ``Attention-based deep multiple instance learning,''
\newblock {\em ArXiv}, vol. abs/1802.04712, 2018.

\bibitem{RAPT:1995}
David Talkin and W~Bastiaan Kleijn,
\newblock ``A robust algorithm for pitch tracking ({RAPT}),''
\newblock {\em Speech coding and synthesis}, vol. 495, pp. 518, 1995.

\bibitem{PoveyASRU2011}
Daniel Povey,
\newblock ``The kaldi speech recognition toolkit,''
\newblock in {\em IEEE 2011 Workshop on Automatic Speech Recognition and
  Understanding}. Dec. 2011, IEEE Signal Processing Society.

\bibitem{waveRNN:2018}
Nal Kalchbrenner, Erich Elsen, Karen Simonyan, Seb Noury, Norman Casagrande,
  Edward Lockhart, Florian Stimberg, Aaron van~den Oord, Sander Dieleman, and
  Koray Kavukcuoglu,
\newblock ``Efficient neural audio synthesis,''
\newblock {\em arXiv preprint arXiv:1802.08435}, 2018.

\end{thebibliography}

\end{document}